\documentclass[a4paper,12pt]{article}
\usepackage{amsfonts}
\usepackage{latexsym}
\usepackage{amssymb}
\usepackage{color}
\date{\empty}

\begin{document}
%\tableofcontents

\title{Discrete Systems in Thermal Physics and Engineering\\ 
--A Glance from Non-Equilibrium Thermodynamics--}
\author{W. Muschik\footnote{Corresponding author:
muschik@physik.tu-berlin.de}
\thanks{In memory of Joseph Kestin}
\\
Institut f\"ur Theoretische Physik\\
Technische Universit\"at Berlin\\
Hardenbergstr. 36\\D-10623 BERLIN,  Germany}
\maketitle

            \newcommand{\be}{\begin{equation}}
            \newcommand{\beg}[1]{\begin{equation}\label{#1}}
            \newcommand{\ee}{\end{equation}\normalsize}
            \newcommand{\bee}[1]{\begin{equation}\label{#1}}
            \newcommand{\bey}{\begin{eqnarray}}
            \newcommand{\byy}[1]{\begin{eqnarray}\label{#1}}
            \newcommand{\eey}{\end{eqnarray}\normalsize}
            \newcommand{\beo}{\begin{eqnarray}\normalsize}
         
            \newcommand{\R}[1]{(\ref{#1})}
            \newcommand{\C}[1]{\cite{#1}}

            \newcommand{\mvec}[1]{\mbox{\boldmath{$#1$}}}
            \newcommand{\x}{(\!\mvec{x}, t)}
            \newcommand{\m}{\mvec{m}}
            \newcommand{\F}{{\cal F}}
            \newcommand{\n}{\mvec{n}}
            \newcommand{\argm}{(\m ,\mvec{x}, t)}
            \newcommand{\argn}{(\n ,\mvec{x}, t)}
            \newcommand{\T}[1]{\widetilde{#1}}
            \newcommand{\U}[1]{\underline{#1}}
            \newcommand{\V}[1]{\overline{#1}}
            \newcommand{\ub}[1]{\underbrace{#1}}
            \newcommand{\X}{\!\mvec{X} (\cdot)}
            \newcommand{\cd}{(\cdot)}
            \newcommand{\Q}{\mbox{\bf Q}}
            \newcommand{\p}{\partial_t}
            \newcommand{\z}{\!\mvec{z}}
            \newcommand{\bu}{\!\mvec{u}}
            \newcommand{\rr}{\!\mvec{r}}
            \newcommand{\w}{\!\mvec{w}}
            \newcommand{\g}{\!\mvec{g}}
            \newcommand{\D}{I\!\!D}
            \newcommand{\se}[1]{_{\mvec{;}#1}}
            \newcommand{\sek}[1]{_{\mvec{;}#1]}}            
            \newcommand{\seb}[1]{_{\mvec{;}#1)}}            
            \newcommand{\ko}[1]{_{\mvec{,}#1}}
            \newcommand{\ab}[1]{_{\mvec{|}#1}}
            \newcommand{\abb}[1]{_{\mvec{||}#1}}
            \newcommand{\td}{{^{\bullet}}}
            \newcommand{\eq}{{_{eq}}}
            \newcommand{\eqo}{{^{eq}}}
            \newcommand{\f}{\varphi}
            \newcommand{\rh}{\varrho}
            \newcommand{\dm}{\diamond\!}
            \newcommand{\seq}{\stackrel{_\bullet}{=}}
            \newcommand{\sta}[2]{\stackrel{_#1}{#2}}
            \newcommand{\om}{\Omega}
            \newcommand{\emp}{\emptyset}
            \newcommand{\bt}{\bowtie}
            \newcommand{\btu}{\boxdot}
            \newcommand{\tup}{_\triangle}
            \newcommand{\tdo}{_\triangledown}
            \newcommand{\Ka}{\frac{\nu^A}{\Theta^A}}
            \newcommand{\K}[1]{\frac{1}{\Theta^{#1}}}
            \newcommand{\ap}{\approx}
            \newcommand{\bg}{\sta{\Box}{=}}
            \newcommand{\si}{\simeq}
            \newcommand{\GG}{\mvec{]}}
            \newcommand{\LL}{\mvec{[}}
            \newcommand{\bx}{\boxtimes}
\newcommand{\Section}[1]{\section{\mbox{}\hspace{-.6cm}.\hspace{.4cm}#1}}
\newcommand{\Subsection}[1]{\subsection{\mbox{}\hspace{-.6cm}.\hspace{.4cm}
\em #1}}

\newcommand{\const}{\textit{const.}}
\newcommand{\vect}[1]{\underline{\ensuremath{#1}}}  %Vektoren
\newcommand{\abl}[2]{\ensuremath{\frac{\partial #1}{\partial #2}}}


\vspace{.6cm}
\noindent
{\bf Abstract} Non-equilibrium processes in Schottky systems generate by projection onto the
equilibrium subspace reversible accompanying processes for which the non-equilibrium variables
are functions of the equilibrium ones. The embedding theorem which guarantees the
compatibility of the accompanying processes with the non-equilibrium entropy is proved.
The non-equilibrium entropy is defined as a state function on the non-equilibrium state space
containing the contact temperature as a non-equilibrium variable. If the entropy production 
does not depend on the internal energy, the contact temperature changes into the thermostatic
temperature also in non-equilibrium, a fact which allows to use temperature as a primitive concept
in non-equilibrium. The dissipation inequality is revisited, and an efficiency of
generalized cyclic processes beyond the Carnot process is achieved.

\section{Preface: Motivation}

This tutorial paper which summarizes some known and some novel results of non-equi\-li\-bri\-um
thermodynamics concerning temperature and entropy is addressed to those who are interested in
fundamentals of thermodynamics of non-equilibrium beyond the usual frame \C{GRMA62}
\C{KJBEJOGR17}. One may be of the opinion that these fundamentals are well known,
but this impression is not true as can be seen by the following argumentation.

Temperature and entropy are well defined in thermostatics, that is the area of reversible "processes"
which do not exist in reality (do not take them for quasi-static processes). But most of the thermal
processes are irreversible: heat conduction, diffusion, friction, chemical reactions.
The question arises: why in all these non-equilibrium items, the equilibrium temperature can be
used successfully, especially since a non-equilibrium entropy is introduced ? There is only one
way to answer this question: a measurable non-equilibrium temperature has to be defined, and one
has to search for conditions which allow to replace this non-equilibrium temperature exactly by the
thermostatic equilibrium temperature (do not think of hypotheses as local equilibrium or
endoreversibily).

In connection with the non-equilibrium entropy, the busy physicists have generated a lot of
"non-equilibrium temperatures" with the shortcoming that these may not fit to the used 
thermometers. Thus it was obvious, first to find a non-equilibrium temperature
fitting to the thermometers and afterwards to define a non-equilibrium entropy. This is the contact
temperature, the starting-point of this paper. Its replacement by the thermostatic temperature
needs some formal, but easy and clear procedures. Of course, several unfamiliar
notations appear, such as embedding theorem, adiabatic uniqueness and thermal stage, but they
all are transparent concepts of non-equilibrium thermodynamics. After some efforts, the elucidation
shows up why in thermal physics the equilibrium temperature can be used exactly even in
non-equilibrium.

\section{Introduction}

There are two kinds of describing thermodynamic phenomena: as a field formulation and as discrete
systems, also called Schottky systems \C{SCHO29}. The field formulation is able to describe more
details than the gross representation by a Schottky system. Both tools have their special realm of
application: field formulation, if one can establish differential equations with corresponding initial
and boundary conditions, and considering a discrete system in interaction with its environmement
\C{MUBE04}, a situation which is frequent in thermal engineering because especially boundary
conditions of possible differential equations are unknown. Here, this second case is considered using 
non-equilibrium thermodynamics of discrete systems.

First of all one has to realize, that non-equilibrium thermodynamics means that a non-equilibrium
entropy and consequently also a non-equilibrium temperature have to be introduced \C{MU18a} and
that these quantities have to be consistent with thermostatics (equilibrium thermodynamics).  
There exists a detailed literature on non-equilibrium temperature \C{CVJ03}, but the problem
arises, if for such a non-equilibrium temperature also a thermometer exits or if this temperature
is only an arithmetical quantity.

There is a huge variety of different thermometers \C{23,1} all measuring "temperature", but
the concept of temperature is first of all only properly defined in equilibrium. For elucidating this fact,
we consider the simple example of a thermometer whose surface\footnote[1]{The thermometer is
here a discrete system which has a volume and a surface, as small as ever.} has different zones of
heat conductivities. Contacted with a non-equilibrium system, the measured "temperature" depends
at the same local position on the orientation of such a "thermometer". Clear is, that this orientation
sensitivity of the thermometer vanishes in equilibrium. A second
example is a thermometer which measures the intensity of radiation which is composed of different
parts of the spectrum. The measured "temperature" depends on the sensitivity distribution of the
thermometer over the spectrum with the result, that different thermometers measure different
"temperatures" at the same object.

For escaping these "thermometer induced" difficulties, a theoretical definition of temperature in
non-equilibrium is helpful as a remedy. Taking the usual arithmetical definitions of temperature $T$
into account \C{1},
\bee{B1}
\mbox{discrete systems:}\quad\frac{1}{T}\ :=\ \frac{\partial S}{\partial U},\qquad
\mbox{field formulation:}\quad\frac{1}{T}\ :=\ \frac{\partial s}{\partial u},
\ee
also these definitions have their malices: First of all, a state space on which $S$ or $s$ are defined
is needed, because the partial derivatives in \R{B1} have no sense without it. Then entropy $S$ or entropy density $s$  and internal
energy $U$ or internal energy density $u$ are needed in equilibrium and out of it. And finally, the
open question is, if there exists a thermometer which measures that temperature $T$. 

To avoid all these uncertainties, a simple idea is the following: why not define a general concept
of temperature which is valid independently of equilibrium or non-equi\-li\-brium and which is
introduced into the theoretical framework by defining the RHSs of \R{B1}, thus determining the
partial derivatives entropy over temperature
\bee{B2}
\frac{1}{\Theta}\ =:\ \frac{\partial S}{\partial U},\qquad
\frac{1}{\Theta}\ =:\ \frac{\partial s}{\partial u}\ ?
\ee
If additionally $\Theta$ is connected with a measuring instruction which "defines" the temperature
$\Theta$ experimentally, temperature comes from the outside into the theoretical framework and
not vice versa \C{MU77,MUBR77,MUBR75}.
Because $\Theta$ will generally be defined independently of equili\-brium or
non-equilibrium, also $S$ and $s$ are according to \R{B2} such generalized quantities.

That these concepts of non-equilibrium temperature and entropy are not only interesting with
respect to theoretical considerations, but also to systems of engineering thermodynamics --discrete
systems-- is demonstrated in the sequel.

The paper is organized as follows: After this introduction, Schottky systems are revi\-sited, state
spaces and processes, the 1st and the 2nd law are considered, and the entropy time rate and
accompanying processes are introduced. Non-equilibrium contact quantities such as temperature,
enthalpy and chemical potentials are defined. After a brief view at internal variables, a
non-equilibrium entropy is established by use of the adiabatic uniqueness and the embedding
theorem. The results are summed up in so-called thermodynamical stages of Schottky systems.
The dissipation inequality and the efficiency of generalized
cyclic processes are treated in the last two sections. The paper finishes with a summary.

\section{Schottky Systems}

Numerous situations in thermal sciences can be described by the interaction of a discrete
system, a Schottky system, with its environment \C{MUBE04,MUBE07}.
Here, the system is in non-equilibrium and the
environment in equilibrium or also in non-equilibrium. For describing such a compound system
consisting of the discrete system and its environment, two kinds of variables are needed: the state
variables describing the states of the sub-systems (discrete system and environment) and the
interaction variables describing the exchanges between the sub-systems \C{MUAS1,MU93}.

A system $\cal G$, described as undecomposed and homogeneous,
which is separated by a partition $\partial {\cal G}$ from its environment
${\cal G}^\Box$ is called a {\em Schottky system} \C{SCHO29}, if the interaction
between $\cal G$ and ${\cal G}^\Box$ through $\partial {\cal G}$ can be described
by\footnote{$^\td$ means "per time", but not neccessarily a time derivative}
\bee{4}
\mbox{heat exchange\ }\sta{\td}{Q},\quad\mbox{power exchange\ }\sta{\td}{W},\quad
\mbox{ and material exchange\ }\sta{\td}{\mvec{n}}\!{^e}.
\ee
The power exchange is related to the work variables $\mvec{a}$ of the system
\bee{5}
\sta{\td}{W}\ =\ {\bf A}\cdot\sta{\td}{\mvec{a}}.
\ee
Here, ${\bf A}$ are the generalized forces which are as well known as the work variables.
Kinetic and potential energy are constant and therefore out of scope. $\sta{\td}{Q}$ is
measurable by calorimetry and the time rate of the mole numbers due to
material exchange $\sta{\td}{\mvec{n}}\!{^e}$ by weigh.

\subsection{State spaces and processes}

A large state space $\cal Z$ \C{MUAS1} is decomposed into its equilibrium subspace
containing the equilibrium variables $\mvec{\Omega}$ and into the non-equilibrium part
spanned by the variables $\mvec{\Xi}$ which describe non-equilibrium and which are not
included in the equilibrium subspace 
\bee{10}
\mvec{Z}\ =\ (\mvec{\Omega},\mvec{\Xi})\ \in\ {\cal Z}.
\ee
The {\em states of equilibrium} $\mvec{\Omega}$ are defined by time independent
states of an isolated Schott\-ky system and are determined by the {\em Zeroth Law}:
The state space of a thermal homogeneous\footnote{that means: there are no adiabatical partitions
dividing the interior of the discrete system into comparments} Schottky system in equiIibrium is
spanned by the work variables $\mvec{a}$, the mole numbers $\mvec{n}$ and the internal
energy $U$
\bee{11}
\mvec{\Omega}\ =\ (\mvec{a},\mvec{n}, U)\quad\longrightarrow\quad
\mvec{Z}\ =\ (\mvec{a},\mvec{n}, U,\mvec{\Xi}).
\ee
A process is defined as a trajectory on the non-equilibrium state space,
\bee{20w}
\mvec{Z}(t)\ =\ \Big(\mvec{a},\mvec{n}, U,\mvec{\Xi}\Big)(t),\quad t=\mbox{time}.
\ee

In equilibrium, the non-equilibrium variables $\mvec{\Xi}$ depend on the equilibrium ones
$\mvec{\Xi}(\mvec{\Omega})$ and \R{11}$_2$ becomes
\bee{21vy}
\mvec{Z}^{eq}\ =\ \Big(\mvec{a},\mvec{n}, U,\mvec{\Xi}(\mvec{a},\mvec{n}, U)\Big).
\ee
This fact gives rise to introduce a projection $\cal P$ of the non-equilibrium state $\mvec{Z}$
onto  the equilibrium subspace \C{MU93} resulting in an equilibrium state $\mvec{\Omega}$
\bee{19w}
{\cal P}\mvec{Z}\ =\ {\cal P}\Big(\mvec{a},\mvec{n}, U,\mvec{\Xi}\Big)\ =\
\mvec{Z}^{eq}\ =\ \Big(\mvec{a},\mvec{n}, U,\mvec{\Xi}(\mvec{a},\mvec{n}, U)\Big)\ =\
\Big(\mvec{\Omega},\mvec{\Xi}(\mvec{\Omega})\Big).
\ee

Because this projection can be performed for each non-equilibrium state of a process \R{20w},
the time is projected onto the equilibrium states\footnote{t is the "slaved time" according to
\R{21v}} now equipped with a time, thus representing an "equilibrium process", a bit strange
denotation because neither a "process" with progress in time takes place on the equilibrium subspace
nor does it exist in nature. Such "processes" are called {\em reversible}
\bee{21v}
{\cal P}\mvec{Z}(t)\ =\
\mvec{Z}^{eq}(t)\ =\ \Big(\mvec{a},\mvec{n}, U,\mvec{\Xi}(\mvec{a},\mvec{n}, U)\Big)(t).
\ee
Consequently, reversible processes are trajectories on the equilibrium subspace generated by a
point-to-point projection of non-equilibrium states onto the equilibrium subspace, thus
keeping the "time" as path parameter of $\mvec{\Omega}(t)$ which is also denoted as
{\em accompanying process} $\mvec{Z}^{eq}(t)$ \C{KE71} which is detailled conssidered in
sect.\ref{ACC}. Although not existing in nature, reversible processes serve as thermostatic
approximation and as mathematical closing of the set of "real" (irreversible) processes which are
defined as trajectories on the non-equilibrium state space.

\subsection{The First Law}

Up to now, the internal energy was introduced in \R{11}$_1$ as one variable of the
equilibrium subspace of a thermally homogeneous Schottky system. The connection
between the time rate of the internal energy of the system and the exchange quantities
through $\partial\cal G$ is establiched by the {\em First Law}
\bee{17}
\sta{\td}{U}\ =\ \sta{\td}{Q} + \mvec{h}\cdot\sta{\td}{\mvec{n}}\!{^e} + \sta{\td}{W} 
\ee
which states that the internal energy $U$ of the system should be conserved in isolated
Schottky systems. The second term of the RHS of \R{17} originates from the fact that the
heat exchange has to be redefined for open systems
($\sta{\td}{\mvec{n}}\!{^e}\neq\mvec{0}$) \C{MUGU99}. Here, $\mvec{h}$ are the
molar enthalpies of the chemical components in $\cal G$. The modified heat exchange
which is combined with the material exchange appearing in the First Law \R{17} was
used by Rolf Haase \C{HA69}.

\subsection{Entropy time rate and Second Law}

Considering a discrete system $\cal G$, a quantity $\bf J$ of ${\cal G}$ is called
{\em balanceable}, if its time rate can be decomposed into an exchange $\bf\Psi$
and a production  $\bf R$
\bee{18}
\sta{\td}{\bf J}\ =\ {\bf\Psi} + {\bf R},\qquad {\bf\Psi}\ =\
{\bf\Phi}+ \varphi\sta{\td}{\mvec n}\!{^e}.
\ee
The exchange is composed of its conductive part $\bf\Phi$ and its convective part
$\varphi\sta{\td}{\mvec n}\!{^e}$.

Doubtless, a non-equilibrium entropy of a Schottky system is a balanceable
quantity\footnote{about the existence of a "non-equilibrium" entropy see \C{MU18a}}.
Presupposing that the power exchange does not contribute to the entropy rate, if the
heat exchange is taken into consideration, we obtain according to \R{4}
\bee{21}
\sta{\td}{S}(\mvec{Z})\ =\ \frac{1}{\Theta}\sta{\td}{Q}+\mvec{s}\cdot\sta{\td}{\mvec n}\!{^e}+\Sigma.
\ee
Here, $\sta{\td}{Q}$ and $\sta{\td}{\mvec n}\!{^e}$ are the exchange quantities,
$\mvec{s}$ the molar entropies, whereas $\Sigma$ is the {\em entropy production} which
is not negative according to the {\em Second Law}
\bee{21a}
\Sigma\ \geq\ 0\ \longrightarrow\
\sta{\td}{S}\ \geq\ \frac{1}{\Theta}\sta{\td}{Q}+\mvec{s}\cdot\sta{\td}{\mvec n}\!{^e}\
\longrightarrow\ 0\ \geq\ \oint\Big[
\frac{1}{\Theta}\sta{\td}{Q}+\mvec{s}\cdot\sta{\td}{\mvec n}\!{^e}\Big]dt.
\ee

State functions on a large state space $\cal Z$ \C{MUAS1} are unique, and 
if the considered Schottky system is adiabatically unique \C{MU09}, the entropy 
rate $\sta{\td}{S}$ is the time derivative of a state space function entropy $S(\mvec{Z})$
according to \R{11}. Consequently, its integration along a cyclic process vanishes, resulting in
Clausius' inequality \R{21a}$_3$.
Because $S$ is a non-equilibrium entropy, also the temperature $\Theta$, the molar entropies
and the entropy production in \R{21} are non-equilibrium quantities\footnote{If $\Theta$ is replaced
in \R{21} by the thermostatic temperature $T^\Box$ of the environment, $\sta{\td}{S}$ is not a
time derivation of a state function because $T^\Box$ does not belong to the Schottky system and
\R{21}$_3$ is not valid.}. The non-equilibrium
temperature $\Theta$ is the contact temperature which is detailled considered in sect.\ref{CONT}.

Presupposing that all chemical components in $\cal G$ have the same temperature
$\Theta$, the non-equilibrium molar entropies $\mvec{s}$ in \R{21} are \C{KES79}
\bee{31z}
\mvec{s}\ =\ \frac{1}{\Theta}\Big(\mvec{h}-\mvec{\mu}\Big),
\ee
with the non-equilibrium molar enthalpies $\mvec{h}$ appearing in the First Law \R{17} and
the non-equilibrium chemical potentials $\mvec{\mu}$. For the present, these non-equilibrium
quantities $(\Theta, \mvec{h}, \mvec{\mu})$ are unknown and only place-holders which will be
determined below in sect.\ref{CONT}.
Consequently taking \R{17} and \R{31z} into account, the entropy time rate \R{21} becomes
\bee{38}
\sta{\td}{S}\ =\ \frac{1}{\Theta}\Big(\sta{\td}{U}
-{\bf A}\cdot\sta{\td}{\mvec{a}}
-\mvec{\mvec{\mu}}\cdot\sta{\td}{\mvec{n}}\!{^e}\Big)+\Sigma.
\ee

Because the external mole number rates $\sta{\td}{\mvec{n}}\!\!{^e}$ are no state
variables\footnote{because they depend on the environment}, but the mole numbers
themselves are included in the equilibrium subspace \R{11}$_1$ according to the
Zeroth Law, the missing term for generating the mole numbers in \R{38} is hidden
in the entropy production
\bee{39}
\Sigma\ =\ -\frac{1}{\Theta}\mvec{\mvec{\mu}}\cdot\sta{\td}{\mvec{n}}\!{^i}
+\Sigma^0,
\ee
and \R{38} results in
\bee{40}
\sta{\td}{S}(\mvec{Z})\ =\ \frac{1}{\Theta}\Big(\sta{\td}{U}
-{\bf A}\cdot\sta{\td}{\mvec{a}}-
\mvec{\mvec{\mu}}\cdot\sta{\td}{\mvec{n}}\Big)+\Sigma^0,\qquad \Sigma^0\ \geq\ 0,
\ee
if the internal mole number changes due to chemical reactions
\bee{18x}
\sta{\td}{\mvec{n}}\!{^i}\ =\ \sta{\td}{\mvec{n}}-\sta{\td}{\mvec{n}}\!{^e}
\ee
are taken into account. Because the bracket in \R{40} contains only rates of equilibrium
variables, those of the non-equilibrium state variables appear in the entropy production
$\Sigma^0$ which is established below in sect.\ref{IECT}.

\subsection{Reversible and accompanying processes}

A reversible "process" is defined as a trajectory on the {\em equilibrium subspace}
$\mvec{\Omega}$ \R{11}$_1$ along that the entropy production \R{39} vanishes by definition
\bee{M0}
\Sigma^*\ =\ -\frac{1}{T^*}\mvec{\mu}^*\cdot\sta{\td}{\mvec{n}}\!{^{i*}}+\Sigma^{0^*}\
\equiv\ 0,\qquad\sta{\td}{\mvec{n}}\!{^{i*}}\ \equiv\ 0,
\ee
because chemical reactions are irreversible processes.
The entropy rate \R{21} of a reversible process is
\bee{M1}
\sta{\td}{S^*}\ :=\ \frac{1}{T^*}\sta{\td}{Q}\!{^*}
+\mvec{s}^*\cdot\sta{\td}{\mvec n}\!^{e*},\qquad\Sigma^*\equiv 0,
\ee
and \R{40} results in
\bee{M2}
\sta{\td}{S^*}\ =\ \frac{1}{T^*}\Big(\sta{\td}{U}\!{^*}
-{\bf A}^*\cdot\sta{\td}{\mvec{a}}\!{^*}-
\mvec{\mu}^*\cdot\sta{\td}{\mvec{n}}\!{^*}\Big),\qquad \Sigma^{0^*}\ \equiv\ 0.
\ee
The First Law of reversible processes is according to \R{17}
\bee{M4}
\sta{\td}{U^*}\ =\ \sta{\td}{Q^*} + \mvec{h}^*\cdot\sta{\td}{\mvec{n}}\!^{e*} + 
\sta{\td}{W}\!{^*}\ =\
\sta{\td}{Q^*}+\mvec{h}^*\cdot\sta{\td}{\mvec{n}}\!{^*}
+{\bf A}^*\cdot\sta{\td}{\mvec{a}}\!{^*},\qquad\sta{\td}{\mvec{n}}\!^{i*}\ \equiv\ 0.
\ee

Special reversible processes are generated by projections \R{19w} of an irreversible process
$\mvec{Z}(t)$ onto the equilibrium subspace: the accompanying processes
${\cal P}\mvec{Z}(t)$\footnote{$\doteq$ marks a setting}, 
\bee{M3}
\sta{\td}{U^*} \doteq\ \sta{\td}{U},\quad\sta{\td}{\mvec{a}}\!{^*} \doteq\ \sta{\td}{\mvec{a}},\quad
\sta{\td}{\mvec{n}}\!{^*} \doteq\ \sta{\td}{\mvec{n}},
\ee
resulting in the First Law and the entropy time rate of accompanying processes
\bee{M2x}
\sta{\td}{U}\ =\
\sta{\td}{Q^*}+\mvec{h}^*\cdot\sta{\td}{\mvec{n}}
+{\bf A}^*\cdot\sta{\td}{\mvec{a}},
\qquad
\sta{\td}{S^*}\!(\mvec{\Omega})\ =\ \frac{1}{T^*}\Big(\sta{\td}{U}
-{\bf A}^*\cdot\sta{\td}{\mvec{a}}-
\mvec{\mu}^*\cdot\sta{\td}{\mvec{n}}\Big).
\ee

The connection between the non-equilibrium quantities 
$(\Theta, \sta{\td}{Q}, \mvec{s}, \mvec{h}, \mvec{\mu}, {\bf A}, \sta{\td}{S})$ and the projected
equilibrium quantities
$(T^*, \sta{\td}{Q^*}, \mvec{s}^*, \mvec{h}^*, \mvec{\mu}^*, {\bf A}^*, \sta{\td}{S^*})$
is investigated below in sect.\ref{ACC}.

\section{Contact Quantities}
\subsection{Defining inequalities}

A discrete non-equilibrium system $\cal G$ is now considered which is surrounded with an equilibrium
reservoir ${\cal G}^\Box$ having a joint surface $\partial{\cal G}\equiv\partial{\cal G}^\Box$,
that means, a compound system ${\cal G}\cup{\cal G}^\Box$ is considered whose sub-systems
have mutual exchanges of heat, power and material. Usually, ${\cal G}^\Box$ is denoted as 
the system's controlling environment. The joint surface represents the partition between the two
sub-systems. Especially, inert partitions are considered which are defined as follows:
An inert partition does not absorb or emit heat, power and material \cite{MU09},
described by the following equations \cite{MUBE04,MUBE07}
\bee{9a}  
\sta{\td}{Q}\ =\ -\sta{\td}{Q}\!{^\Box},\qquad
\sta{\td}{W}\ = \ {\bf A}\cdot\sta{\td}{\mvec{a}}\ =\ 
{\bf A}^\Box\cdot\sta{\td}{\mvec{a}}\ = \ -\sta{\td}{W}\!{^\Box},\qquad
\sta{\td}{\mvec{n}}\!{^e}\ =\ -\sta{\td}{\mvec{n}}\!{^{\Box e}}.
\ee
The $^\Box$-quantities belong to the system's controlling environment ${\cal G}^\Box$.
The power done on the system is performed by the environment using its generalised
forces ${\bf A}^\Box$ and orientated at the work variables of the system. The permeability of
$\partial\cal G$ to heat, power and material is described by \R{9a}. 

Now, special compound systems are considered which satisfy the following
\begin{center}
\parbox[b]{12.5cm}{ 
{\sf Axiom:}The non-negative entropy time rate
of the isolated compound system ${\cal G}^\Box \cup {\cal G}$ is the sum of the entropy
time rates of the sub-systems.  
}
\end{center}
Taking \R{9a}$_{1,3}$ into account, the total entropy time rate of the isolated compound system
${\cal G}\cup{\cal G}^\Box$ is according to this axiom
\bey\nonumber
\sta{\td}{S}\!^{tot}\ :=\ \sta{\td}{S} + \sta{\td}{S}\!{^\Box} &=&
\frac{1}{\Theta}\sta{\td}{Q}+\mvec{s}\cdot\sta{\td}{\mvec n}\!{^e}+
\frac{1}{T^\Box}\sta{\td}{Q}\!{^\Box}+
\mvec{s}{^\Box}\cdot\sta{\td}{\mvec n}\!^{\Box e}+\Sigma\ =\
\\ \label{25}
&=&\Big(\frac{1}{\Theta}-\frac{1}{T^\Box}\Big)\sta{\td}{Q}
+(\mvec{s}-\mvec{s}{^\Box})\cdot\sta{\td}{\mvec n}\!{^e}+\Sigma\ \geq\ 0. 
\eey

Up to now, $\Theta$ and $\mvec{s}$ are place holders in the dissipation inequality \R{25}$_2$
for the unknown contact quantities temperature and molar entropy.
Whereas $\Sigma$ is the internal entropy production of the sub-system ${\cal G}$ 
according to \R{21}, $(1/\Theta-1/T^\Box)\sta{\td}{Q}$ and $(\mvec{s}-\mvec{s}{^\Box})\cdot\sta{\td}{\mvec n}\!{^e}$ represent the entropy production of the heat
and material exchanges between the sub-systems of the compound system. If the system is a
reversible one ($\Sigma=0$), these exchanges have to be compatible with the dissipation inequality
\R{25}$_2$. Because heat and material exchanges are independent of each other, the following
{\em defining inequalities} \C{MU18a,MU09}
\bee{25a}
\Big(\frac{1}{\Theta}-\frac{1}{T^\Box}\Big)\sta{\td}{Q}\ \sta{*}{\geq}\ 0\qquad
(\mvec{s}-\mvec{s}{^\Box})\cdot\sta{\td}{\mvec n}\!{^e}\ \sta{*}{\geq}\ \mvec{0}
\ee
are demanded for defining the place holders {\em contact temperature} $\Theta$ and 
{\em non-equilibrium molar entropy} $\mvec{s}$ which now are ascribed to
the sub-system $\cal G$ of the compound system ${\cal G}^\Box\cup{\cal G}$
\C{MU77,MUBR77,MUBR75,MU09}.

In more detail: the non-equilibrium Schottky systen $\cal G$ is in contact with a
thermally homogeneous equilibrium environment ${\cal G}^\Box$ of the thermostatic
temperature $T^\Box$ and the molar equilibrium entropies $\mvec{s}^\Box$.  All chemical
components in ${\cal G}^\Box$ have the same temperature $T^\Box$. Then the inequalities
\R{25a} define the contact temperature $\Theta$ and the molar entropies $\mvec{s}$ as is
demonstrated in the next section.

\subsection{Contact temperature, neq-enthalpy and chemical potential\label{CONT}}

Taking \R{31z} into account for equilibrium and non-equilibrium processes, the defining inequality
\R{25a}$_2$ of the molar entropies can be satisfied, if the additional inequalities for the molar
enthalpies and for the chemical potentials are demanded
\bee{29}
\Big(\frac{\mvec{h}}{\Theta}-\frac{\mvec{h}^\Box}{T^\Box}\Big)\cdot\sta{\td}{\mvec n}\!{^e}\ 
\sta{*}{\geq}\ 0,\qquad
\Big(\frac{\mvec{\mu}^\Box}{T^\Box}
-\frac{\mvec{\mu}}{\Theta}\Big)\cdot\sta{\td}{\mvec n}\!{^e}\ \sta{*}{\geq}\ 0.
\ee

Because the system's environment ${\cal G}^\Box$ is an equilibrium system, the quantities
$\boxplus^\Box$ are well known. As already performed in the introduction, also the non-equilibrium
quantities --temperature, molar enthalpies and chemical potentials-- should be measurable
quantities and not only being defined by arithmetical items. This measurability is guaranteed by
special chosen environments ${\cal G}^\Box$ of $\cal G$, as demonstrated below.

Special equilibrium environments are considered which cause that certain
non-equilibrium contact rates vanish
\byy{29a}
&&{\cal G}^\Box_\odot :\quad \sta{\td}{Q}\!_\odot\ =\ 0,\longrightarrow\  \Theta\ =\ T^\Box_\odot,\quad\mbox{according to \R{25a}$_1$,}
\\ \label{29b}
&&{\cal G}^\Box_{nj}:\quad \sta{\td}{n_j}\ =\ 0,\longrightarrow\ 
h_j\ =\ \frac{\Theta}{T^\Box}h^\Box_j,\quad
\mu_j\ =\ \frac{\Theta}{T^\Box}\mu^\Box_j ,\quad\mbox{according to \R{29}.}
\eey
For deriving \R{29a}$_2$ and \R{29b}$_{2,3}$ from \R{25a}$_1$ and \R{29}, the following
proposition \C{MU84} is used:
\bee{30}
f({\bf X})\cdot {\bf X}\ \geq\ {\bf 0}\ (\mbox{for all}\ {\bf X}\wedge f \ \mbox{continuous at}\ 
{\bf X} = {\bf 0})\ \Longrightarrow\ f({\bf 0}) = {\bf 0}.
\ee
Consequently, the non-equilibrium contact quantities {\em contact temperature}
$\Theta$, {\em molar non-equilibrium enthalpies} $\mvec{h}$ and
{\em non-equilibrium chemical potentials} $\mvec{\mu}$ are defined by equilibrium
quantities of special contacting equilibrium environments according to \R{29a} and
\R{29b}. In more detail:

According to \R{29a}, the following definition is
made \cite{MU77,MUBR77,BRMU75}
\begin{center}
\parbox[b]{12.5cm}{
{\sf Definition:} The system's contact temperature $\Theta$ is that thermostatic temperature
$T^\Box_\odot$ of the system's equilibrium environment for which the net heat exchange
$\sta{\td}{Q}\!_\odot$ between the system and this environment through an inert partition vanishes
by change of sign.
}
\end{center}
Corresponding definitions are valid for the non-equilibrium molar enthalpies and chemical potentials
according to \R{29b}.

\subsection{The extended Clausius inequality}

Presupposing that the non-equilibrium entropy $S(\mvec{Z})$ is a state function on the
non-equilibrium state space \R{11}$_2$\footnote{more details in sect.\ref{NEE}}, the inequality
\R{21a}$_3$ is valid
\bee{21ax}
 0\ \geq\ \oint\Big[
\frac{1}{\Theta}\sta{\td}{Q}+\mvec{s}\cdot\sta{\td}{\mvec n}\!{^e}\Big]dt.
\ee
Using the defining inequalities \R{25a}
\bee{30a}
\frac{1}{\Theta}\sta{\td}{Q}\ \geq\ \frac{1}{T^\Box}\sta{\td}{Q},\qquad
\mvec{s}\cdot\sta{\td}{\mvec n}\!{^e}\ \geq\ \mvec{s}^\Box\cdot\sta{\td}{\mvec n}\!{^e},
\ee
\R{21ax} yields the extended Clausius inequality \R{30b}$_1$  and in the usual one \R{30b}$_2$
\bee{30b}
 0\ \geq\ \oint\Big[
\frac{1}{\Theta}\sta{\td}{Q}+\mvec{s}\cdot\sta{\td}{\mvec n}\!{^e}\Big]dt\ \geq\
 \oint\Big[
\frac{1}{T^\Box}\sta{\td}{Q}+\mvec{s}^\Box\cdot\sta{\td}{\mvec n}\!{^e}\Big]dt,
\ee
resulting in the
\begin{center}
\parbox[b]{12.5cm}{
{\sf Proposition:} If the non-equilibrium entropy is a state function on the non-equilibrium state
space $(\mvec{a},\mvec{n},U,\Theta,\mvec{\xi})$, and if the place-holders $\Theta$ and
$\mvec{s}$ are identified as contact temperature and non-equilibrium molar entropies by the
defining inequalities \R{30a}, the inequality \R{21ax} changes into the extended Clausius
inequality by which the usual Clausius inequality \R{30b}$_2$ can be proved.
}
\end{center}
Besides the exchange quantities $\sta{\td}{Q}$ and $\sta{\td}{\mvec n}\!{^e}$, the extended
Clausius inequality contains quantities which belong to the Schottky system --$\Theta$ and
$\mvec{s}$-- which are replaced by quantities of the environment --$T^\Box$ and
$\mvec{s}^\Box$-- in the usual Clausius inequality.

\subsection{Internal energy and contact temperature\label{IECT}}

As easily to demonstrate, contact temperature $\Theta$ and internal energy $U$ are
independent of each other in non-equilibrium. For this purpose, a rigid partition
$\partial\cal G$ ($\sta{\td}{\mvec{a}} \equiv\mvec{0}$) between the Schottky system
$\cal G$ and its equilibrium environment ${\cal G}^\Box$ is chosen which is impervious
to matter $(\sta{\td}{\mvec{n}}\!{^e}\equiv 0)$ and a time-dependent environment
temperature $T^\Box(t)$ which is always set equal to the value of the momentary
contact temperature $\Theta(t)$ of $\cal G$, resulting according to \R{25a} and
\R{17} in
\bee{33}
T^\Box (t)\sta{*}{=}\Theta (t)\ \longrightarrow\ \sta{\td}{Q}\ =\ 0\ 
\longrightarrow\ \sta{\td}{U}\ =\ 0.
\ee
Because $\Theta$ is time-dependent and $U$ is constant, totally different from
thermostatics, both quantities are independent of each other.

Because the contact temperature is independent of the internal energy, it represents an
additional variable in \R{40} which is included in $\Sigma^0$. The choice of further
non-equilibrium variables depends on the system in consideration. Here,
{\em internal variables} $\mvec{\xi}$ are chosen because they allow a great flexibility
of describing non-equilibria \C{MU90,MUMAU94}. Consequently, the created
non-equilibrium state space and the entropy production caused by the contact
temperature and the internal variables are
\bee{41}
\mvec{Z}\ =\ (\mvec{a},\mvec{n},U,\Theta,\mvec{\xi}),
\qquad \Sigma^0 =
\alpha\sta{\td}{\Theta}+\mvec{\beta}\cdot\sta{\td}{\mvec{\xi}}\ \geq\ 0.
\ee
According to \R{19w}, the corresponding equilibrium subspace is spanned by
\bee{M5}
{\cal P}\mvec{Z}\ =\ 
{\cal P}\Big(\mvec{a},\mvec{n}, U, \Theta,\mvec{\xi}\Big) =
\Big(\mvec{a},\mvec{n}, U, \Theta(\mvec{a},\mvec{n}, U),
\mvec{\xi}(\mvec{a},\mvec{n}, U)\Big) =
\Big(\mvec{\Omega},\Theta(\mvec{\Omega}),\mvec{\xi}(\mvec{\Omega})\Big).
\ee
By use of these state spaces in connection with the entropy production generated by the contact
temperature and the internal variables, different stages of thermodynamics are introduced
below in sect.\ref{STTH}. But first of all, some remarks on internal variables are necessary.

\section{Brief View at Internal Variables\label{IV} }

Historically, the concept of internal variables can be traced back to Bridgman \C{16}, Meixner
\C{19}, Maugin \C{11z} and many others. The introduction of internal variables makes possible
to use large state spaces, that means,
material properties can be described by mappings defined on the state space variables $\mvec{Z}$,
thus avoiding the use of their histories which appear in small state spaces \C{MUAS1,MU93,MU96}.
Consequently, internal variables allow to use the methods of Irreversible and/or Extended
Thermodynamics \C{7}.

Internal variables cannot be chosen arbitrarily: there are seven concepts which restrict their
introduction \C{MU90}. The most essential ones are:\newline
(i) Internal variables need a model or an interpretation,\newline
(ii) Beyond the constitutive and balance equations, internal variables require rate
equations which can be adapted to different situations, making the use of internal variables flexible
and versatile,\newline
(iii) The time rates of the internal variables do not occur in the work differential of the First Law,
\newline
(iv) An isolation of the discrete system does not influence the internal variables,\newline
(v) In equilibrium, the internal variables become dependent on the variables of the equili\-brium
subspace, if the equilibrium is unconstraint.\newline
Satisfying these concepts, the internal variables entertain an ambiguous relationship with
microstructure \C{MUMAU94}.

As the last term of \R{41} shows, internal variables must be complemented by an evolution
law\footnote[12]{\C{7}, 3.5, 4.7.B, \C{MUMAU94}I, 4.} which may have the shape
\bee{B25}
\sta{\td}{\mvec{\xi}}\ =\ \mvec{f}(U,\mvec{a},\mvec{\xi})
+ \mvec{g}(U,\mvec{a},\mvec{\xi})\sta{\td}{U}
+\  {\bf h}(U,\mvec{a},\mvec{\xi})\cdot\sta{\td}{\mvec{a}}.
\ee
Special one-dimensional cases are
\byy{B26}
\mbox{relaxation type:}&&\quad \sta{\td}{\xi}(t)\ =\ -\frac{1}{\tau(U,\mvec{a},\Theta)}\Big(\xi(t)
-\xi^{eq}\Big),
\\ \label{B27}
\mbox{reaction type \C{MU90}:}&&\quad  \sta{\td}{\xi}(t)\ =\ \gamma(U,\mvec{a},\Theta)\Big[
1-\exp\Big(-\mu(t)\beta(U,\mvec{a},\Theta)\Big)\Big].
\eey

Clear is that the contact temperature $\Theta$ and $\mvec{\xi}$ are different types of internal
variables \C{MU18}: $\sta{\td}{\Theta}$ is entropy generating because the contact temperature is
indepedent of the internal energy, and $\sta{\td}{\mvec{\xi}}$ is entropy producing due to
irreversible processes in the system ${\cal G}$. This difference allows to introduce a classification
into stages of thermodynamics below in sect.\ref{STTH}.

\section{Non-Equilibrium Entropy\label{NEE}}
\subsection{Adiabatical Uniqueness}

Up to now, the projection of a non-equilibrium state $\mvec{Z}$ onto an equilibrium one
${\cal P}\mvec{Z}$ was formally introduced according to \R{M5}. Now, the physical
meaning of ${\cal P}$ is demonstrated in two steps: Consider an arbitrary non-equilibrium state
marked by $C$ 
\bee{M13a}
\mvec{Z}_C\ =\ (U_C,\mvec{a}_C,\mvec{n}_C,\Theta_C,\mvec{\xi}_C)
\ee
and two different processes both starting at $\mvec{Z}_C$
\bey\nonumber
1)&&\mbox{in an isolated system without chemical reactions}
\\ \label{M13b}
&&{\cal I}: \quad \sta{\td}{U}\ =\ 0,\ \sta{\td}{\mvec{a}}\ =\ \mvec{0},\
\sta{\td}{\mvec{n}}\ =\ \mvec{0},
\\ \nonumber
2)&&\mbox{in an open system with chemical reactions}
\\ \label{M13c}
&&{\cal T}: \quad \sta{\td}{U}\ =\ 0,\ \sta{\td}{\mvec{a}}\ =\ \mvec{0},\
\sta{\td}{\mvec{n}}\!{^e}\ =\ -\sta{\td}{\mvec{n}}\!{^i}\ 
\ \longrightarrow\ \sta{\td}{\mvec{n}}\ =\ \mvec{0},
\\ \label{M13d}
&&\hspace{1cm}\sta{\td}{U}\ =\ \sta{\td}{Q}
+\mvec{h}\cdot\sta{\td}{\mvec{n}}\!{^e}+\sta{\td}{W}\
\ \longrightarrow\ \sta{\td}{Q}\ =\ \mvec{h}\cdot\sta{\td}{\mvec{n}}\!{^i}.
\eey

Now, the process \#1 is considered: Presupposing that the isolated system is adia\-ba\-ti\-cally
unique \C{MU09}
\begin{center}
\parbox[b]{12.5cm}{
{\sf Definition:} A Schottky system is called {\em adiabatically unique},
if for each arbitrary, but fixed
non-equilibrium state $C$ after isolation of the system 
the relaxation process ends always in the same final equilibrium state $Aeq$, 
independently of how the process into $C$ was performed.}
\end{center}
The entropy difference along ${\cal I}$ is according to \R{40}$_1$, \R{41}$_2$ and \R{M13b}, 
\bey\nonumber
{\cal I}\int_C^{Aeq}\sta{\td}{S}(\mvec{Z})dt =
{\cal I}\int_C^{Aeq}\Big(\alpha\sta{\td}{\Theta}+\mvec{\beta}\cdot\sta{\td}{\mvec{\xi}}\Big)dt
 &=& S({\cal P}\mvec{Z}_C)-S(\mvec{Z}_C)
\\ \label{M14}
 &=& S^*(\mvec{\Omega}_{Aeq})-S(\mvec{Z}_C).
\eey
Because of the same entropy production according to \R{M13c} and \R{M13d}, the process \#2
along ${\cal T}$ results likewise in the same final equilibrium state $Aeq$ although the process runs
in an open system with chemical reactions
\bee{M15}
{\cal T}\int_C^{Aeq}\sta{\td}{S}(\mvec{Z})dt\ =\
{\cal T}\int_C^{Aeq}\Big(\alpha\sta{\td}{\Theta}+\mvec{\beta}\cdot\sta{\td}{\mvec{\xi}}\Big)dt
=\ S^*(\mvec{\Omega}_{Aeq})-S(\mvec{Z}_C).
\ee
Due to the adiabatical uniqueness, $A_{eq}$ is clearly accessible from
$\mvec{Z}_C$ and the entropy difference, \R{M14}$_2$ and \R{M15}$_2$, between them is
independent of the system's closing. According to the projection \R{M5}, the equilibrium variables
of the non-equilibrium state \R{M13a} are tranferred to the equilibrium state
\bee{M16}
S({\cal P}\mvec{Z}_C)\ =\
S\Big(U_C,\mvec{a}_C,\mvec{n}_C,\Theta(U_C,\mvec{a}_C,\mvec{n}_C),\mvec{\xi}(U_C,\mvec{a}_C,\mvec{n}_C)\Big).
\ee

Vice versa, the equation \R{M14}$_2$ can be interpreted as a definition of a non-equilibrium entropy
\C{MU93}
\bee{M16x}
S(\mvec{Z}_C)\ :=\ S^*(\mvec{\Omega}_{Aeq})-{\cal I}\int_C^{Aeq}
\Big(\alpha\sta{\td}{\Theta}+\mvec{\beta}\cdot\sta{\td}{\mvec{\xi}}\Big)dt,
\ee
if the equilibrium entropy and the entropy production are known.

\subsection{The Embedding Theorem}

Now the entropy time rates are considered which follow from \R{M16x}, that means, the
non-equilibrium state $\mvec{Z}(\tau)$ and its projection onto the equilibrium subspace 
${\cal P}\mvec{Z}(\tau)$ depend on the time $\tau$. If this time differentiation
\bee{M16a}
\frac{d}{d\tau}\boxplus\ \equiv\ \sta{\blacktriangledown}{\boxplus}
\ee
is introduced, \R{M16x} results in
\bee{16b}
\sta{\blacktriangledown}{S}(\mvec{Z})(\tau)\ =\ 
\sta{\blacktriangledown}{S}({\cal P}\mvec{Z})(\tau)
-\frac{d}{d\tau}\Big[{\cal I}\int_{{\bf Z}(\tau)}^{{\cal P}{\bf Z}(\tau)}
\Big(\alpha\sta{\td}{\Theta}+\mvec{\beta}\cdot\sta{\td}{\mvec{\xi}}\Big)dt\Big].
\ee
The entropy difference between two equilibrium states $Aeq$ and $Beq$ is obtained by integration
along an arbitrary process $S$ on the non-equilibrium subspace
\bey\nonumber
{\cal S}\int_{Aeq}^{Beq}\sta{\blacktriangledown}{S}(\mvec{Z})(\tau)d\tau\ =\hspace{10cm}
\\ \label{16c} 
=\ {\cal R}\int_{Aeq}^{Beq}\sta{\blacktriangledown}{S}({\cal P}\mvec{Z})(\tau)d\tau
-\int_{Aeq}^{Beq}\Big\{
\frac{d}{d\tau}\Big[{\cal I}\int_{{\bf Z}(\tau)}^{{\cal P}{\bf Z}(\tau)}
\Big(\alpha\sta{\td}{\Theta}+\mvec{\beta}\cdot\sta{\td}{\mvec{\xi}}\Big)dt\Big]\Big\}d\tau.
\eey
Here, $\cal R$ is the reversible accompanying process of $\cal S$.
The last term of \R{16c} results in
\bee{16d}
\Big[{\cal I}\int_{{\bf Z}(\tau)}^{{\cal P}{\bf Z}(\tau)}
\Big(\alpha\sta{\td}{\Theta}+\mvec{\beta}\cdot\sta{\td}{\mvec{\xi}}\Big)dt\Big]\Big|^{Beq}_{Aeq}
\ =\ 0,\quad\mbox{because: }{\cal P}\mvec{Z}_{Beq}=\mvec{Z}_{Beq},\
{\cal P}\mvec{Z}_{Aeq}=\mvec{Z}_{Aeq}.
\ee
Consequently, the {\em embedding theorem} is proved
\bee{16e}
{\cal S}\int_{Aeq}^{Beq}\sta{\td}{S}(\mvec{Z})(t)dt\ 
=\ {\cal R}\int_{Aeq}^{Beq}\sta{\td}{S}({\cal P}\mvec{Z})(t)dt\ =\ S^*_{Beq}-S^*_{Aeq},
\ee
stating that the time rate of a non-equilibrium entropy $\sta{\td}{S}(\mvec{Z})$ integrated
along an irreversible process ${\cal S}$ between two equilibrium states $Aeq$ and $Beq$
is equal to the integral of the "time rate" of the equilibrium entropy
$\sta{\td}{S}({\cal P}\mvec{Z})$
along the reversible accompanying process ${\cal R}$. The embedding theorem ensures that the
non-equilibrium entropy is compa\-tible with the corresponding equilibrium entropy, that means, 
non-equilibrium entropies have to be defined with respect to the accompanying equilibrium ones, as
done by the definition \R{M16x}.

\subsection{Accompanying entropy time rate\label{ACC}}

Starting with the reversible "process" \R{M5}
\bee{M6}
{\cal P}\mvec{Z}(t)\ =\ 
{\cal P}\Big(\mvec{a},\mvec{n}, U, \Theta,\mvec{\xi}\Big)(t)\ =\
\Big(\mvec{a},\mvec{n}, U, \Theta(\mvec{a},\mvec{n}, U), 
\mvec{\xi}(\mvec{a},\mvec{n}, U)\Big)(t)\ =\ \Big(\mvec{\Omega}\Big)(t)
\ee
which is the {\em accompanying process} belonging to $\mvec{Z}(t)=
\Big(\mvec{a},\mvec{n}, U, \Theta,\mvec{\xi}\Big)(t)$,
the corresponding entropy time rates are according to \R{40}, \R{41}$_2$ and \R{M6}$_2$
\byy{M6a}
\sta{\td}{S}(\mvec{Z}) &=&
 \frac{1}{\Theta}\Big(\sta{\td}{U}
-{\bf A}(\mvec{Z})\cdot\sta{\td}{\mvec{a}}-
\mvec{\mu}(\mvec{Z})\cdot\sta{\td}{\mvec{n}}\Big)
+\alpha(\mvec{Z})\sta{\td}{\Theta}+\mvec{\beta}(\mvec{Z})\cdot\sta{\td}{\mvec{\xi}},
\\ \nonumber
\sta{\td}{S}({\cal P}\mvec{Z}) &=&
\frac{1}{\Theta(\mvec{\Omega})}\Big(\sta{\td}{U}
-{\bf A}(\mvec{\Omega})\cdot\sta{\td}{\mvec{a}}-
\mvec{\mu}(\mvec{\Omega})\cdot\sta{\td}{\mvec{n}}\Big)
+\alpha(\mvec{\Omega})\sta{\td}{\Theta}(\mvec{\Omega})
+\mvec{\beta}(\mvec{\Omega})\cdot\sta{\td}{\mvec{\xi}}(\mvec{\Omega})\ =
\\ \nonumber
&=&\Big(\frac{1}{\Theta(\mvec{\Omega})}+ \alpha(\mvec{\Omega})\frac{\partial \Theta}{\partial U}
+\mvec{\beta}(\mvec{\Omega})\cdot\frac{\partial \mvec{\xi}}{\partial U}\Big)\sta{\td}{U}-
\\ \nonumber
&&-\Big(\frac{\bf A(\mvec{\Omega})}{\Theta(\mvec{\Omega})}-\alpha(\mvec{\Omega})\frac{\partial \Theta}{\partial \mvec{a}}
-\mvec{\beta}(\mvec{\Omega})\cdot\frac{\partial \mvec{\xi}}{\partial \mvec{a}}\Big)\cdot\sta{\td}{\mvec{a}}-
\\ \nonumber
&&-\Big(\frac{\mvec{\mu}(\mvec{\Omega})}{\Theta(\mvec{\Omega})}-\alpha(\mvec{\Omega})\frac{\partial \Theta}{\partial \mvec{n}}
-\mvec{\beta}(\mvec{\Omega})\cdot\frac{\partial \mvec{\xi}}{\partial \mvec{n}}\Big)\cdot\sta{\td}{\mvec{n}}\
\equiv\ 
\\ \label{M6b}
&&\equiv\
\sta{\td}{S}\!{^*}(\mvec{\Omega})\ =\ \frac{1}{T^*}\Big(\sta{\td}{U}
-{\bf A}^*\cdot\sta{\td}{\mvec{a}}-
\mvec{\mu}^*\cdot\sta{\td}{\mvec{n}}\Big),\hspace{.9cm}
\eey 
resulting in
\byy{M9}
\frac{1}{T^*}&=& \frac{1}{\Theta(\mvec{\Omega})}+ \alpha(\mvec{\Omega})\frac{\partial \Theta}{\partial U}
+\mvec{\beta}(\mvec{\Omega})\cdot\frac{\partial \mvec{\xi}}{\partial U},\quad|\sta{\td}{U}
\\ \label{M10}
\frac{{\bf A}^*}{T^*}&=&\frac{\bf A(\mvec{\Omega})}{\Theta(\mvec{\Omega})}
-\alpha(\mvec{\Omega})\frac{\partial \Theta}{\partial \mvec{a}}
-\mvec{\beta}(\mvec{\Omega})\cdot\frac{\partial \mvec{\xi}}{\partial \mvec{a}},
\quad|-\sta{\td}{\mvec{a}}
\\ \label{M11}
\frac{\mu^*}{T^*}&=&\frac{\mvec{\mu}(\mvec{\Omega})}{\Theta(\mvec{\Omega})}
-\alpha(\mvec{\Omega})\frac{\partial \Theta}{\partial \mvec{n}}
-\mvec{\beta}(\mvec{\Omega})\cdot\frac{\partial \mvec{\xi}}{\partial \mvec{n}},
\quad|-\sta{\td}{\mvec{n}}.
\eey

The entropy time rate of the original irreversible process is \R{M6a}, that of the accompanying
process
generated by projection \R{M6} is \R{M6b}. The entropy rate of the accompanying process \R{M6b}
has two different, but equivalent representations: one defined on the space ${\cal P}\mvec{Z}$, 
the other one is defined on the equilibrium subspace $\mvec{\Omega}$.
Multiplying \R{M9} to \R{M11} as specified and summing up, results as expected in \R{M6b}$_3$
\bee{M12}
\sta{\td}{S}{^*}(\mvec{\Omega})\ =\ \sta{\td}{S}({\cal P}\mvec{Z})
\ee
and demonstrates that the entropy production of an accompanying process vanishes.

\section{Thermodynamical Stages of Schottky Systems\label{STTH}}
\subsection{The stage of contact quantities}

The general case is characterized by processes on the non-equilibrium state space \R{41}$_1$
%$\mvec{Z}(t)=\Big(\mvec{a},\mvec{n}, U, \Theta,\mvec{\xi}\Big)(t)$ 
for which contact temperature and internal energy are independent state variables.
The time rate of the non-equilibrium entropy \R{M6a} results in the integrability conditions
\byy{M18}
\frac{\partial S(\mvec{Z})}{\partial U}\ =\ \frac{1}{\Theta},\quad
\frac{\partial S(\mvec{Z})}{\partial \mvec{a}}\ =\ -\frac{\mvec{A}}{\Theta},\quad
\frac{\partial S(\mvec{Z})}{\partial \mvec{n}}\ =\ -\frac{\mvec{\mu}}{\Theta}, \\
\label{M19}
\frac{\partial S(\mvec{Z})}{\partial \Theta}\ =\ \alpha ,\quad
\frac{\partial S(\mvec{Z})}{\partial \mvec{\xi}}\ =\ \mvec{\beta}. 
\eey
Because $\Theta$ and $U$ are independent of each other according to \R{33}, \R{M18}$_1$ can
be integrated immediately
\bee{M19a}
S(U,\mvec{a},\mvec{n},\Theta , \mvec{\xi})\ =\ 
\frac{1}{\Theta}U + K(\mvec{a},\mvec{n},\Theta , \mvec{\xi}).
\ee
Consequently, the non-equilibrium entropy is a linear function of the internal
energy. Here
\bee{M20}
-\Theta K\ =\ F(\mvec{a},\mvec{n},\Theta , \mvec{\xi})
\ee
is the free energy $F$. From the integrability conditions \R{M18} and \R{M19} follows that except of
$\alpha$, all constitutive equations do not depend on the internal 
energy $U$, but instead on the contact temperature:
\byy{48y}
\frac{\partial}{\partial\mvec{a}}\frac{\partial S}{\partial U}&=& \mvec{0}
\quad\Longrightarrow\quad \frac{\partial\mvec{A}}{\partial U}\ =\ \mvec{0},
\\ \label{49y}
\frac{\partial}{\partial\mvec{n}}\frac{\partial S}{\partial U} &=& \mvec{0}
\quad\Longrightarrow\quad \frac{\partial\mvec{\mu}}{\partial U}\ =\ \mvec{0},
\\ \label{50y}
\frac{\partial}{\partial\mvec{\xi}}\frac{\partial S}{\partial U}&=& \mvec{0}
\quad\Longrightarrow\quad \frac{\partial\mvec{\beta}}{\partial U}\ =\ \mvec{0},
\\ \label{51y}
\frac{\partial}{\partial\Theta}\frac{\partial S}{\partial U}&=&
-\frac{1}{\Theta ^2}\ =\  
\frac{\partial\alpha}{\partial U}\quad\Longrightarrow\quad\alpha\ =\ -\frac{U}{\Theta^2}
+ L(\mvec{a},\mvec{n},\Theta , \mvec{\xi}).
\eey
Differentiating \R{M19a} and comparison with \R{M18} and \R{M19} results for 
the free energy \R{M20} in
\bee{53y}
\frac{\partial F}{\partial\Theta}\ =\ -S - \Theta\alpha,\quad
\frac{\partial F}{\partial\mvec{a}}\ =\ \mvec{A},\quad
\frac{\partial F}{\partial\mvec{n}}\ =\ \mvec{\mu},\quad
\frac{\partial F}{\partial\mvec{\xi}}\ =\ -\Theta\mvec{\beta}.
\ee
According to \R{51y}$_3$, the constitutive mapping $\alpha$ is
linear in the internal energy as well as the non-equilibrium entropy \R{M19a}.

\subsection{The endoreversible stage}

The endoreversible case is characterized by "processes" on the equilibrium subspace \R{M5}
which are projections of real running processes
\R{M6a}. For these reversible accompanying processes, the contact
temperature and the internal variables depend on the equilibrium variables.
Consequently, the (contact) temperature depends on the internal energy. Because of the
endoreversibility, there are no internal variables. This situation corresponds to the hypothesis
of local equilibrium in the field formulation of thermodynamics.

The time rate of the entropy \R{M6b}$_4$
\bee{50yy}
\sta{\td}{S}\!{^*}(\mvec{\Omega})\ =\ \frac{1}{T^*}\Big(\sta{\td}{U}
-{\bf A}^*\cdot\sta{\td}{\mvec{a}}-
\mvec{\mvec{\mu}}^*\cdot\sta{\td}{\mvec{n}}\Big),\qquad
{\bf A}^*:={\bf A}(\mvec{\Omega}),\ \mvec{\mu}^*:=\mvec{\mu}(\mvec{\Omega})
\ee
results in the usual integrability conditions
\bee{M21}
\frac{\partial S(\mvec{\Omega})}{\partial U}\ =\ \frac{1}{T^*},\quad
\frac{\partial S(\mvec{\Omega})}{\partial\mvec{a}}\ =\ -\frac{\mvec{A}^*}{T^*},\quad
\frac{\partial S(\mvec{\Omega})}{\partial\mvec{n}}\ =\ -\frac{\mvec{\mu}^*}{T^*}
\ee
whose RHSs are given by \R{M9} to \R{M11}, if the stage of contact quantities is taken as a
background.
 
The entropy production along ${\cal P}\mvec{Z}(t)$ vanishes, the considered Schottky system
undergoes a reversible accompanying process. Entropy production occurs, if different such Schottky
systems interact with each other: the endoreversible case of Finite Time Thermodynamics
\C{MUHO20}.

Because there exist irreversible processes making use of a temperature which depends on the
internal energy, a third stage of thermodynamics has to be formulated in the next section.

\subsection{The stage of Thermal Physics and Engineering}

Considering the independence of the contact temperature $\Theta$ of the internal energy $U$ as
treated in sect.\ref{IECT}, the entropy time rate of an irreversible process is given by \R{M6a} and
that of its (reversible) accompanying process by \R{M6b}. The entropy itself has the shape \R{M19a}.
Especially in thermal engineering, a temperature is used which is connected with the internal energy
and which is different from the thermostatic temperature \R{M9} of the accompanying process.
A projection ${\cal P}^+$ which suppresses the independence of the contact temperature of the
internal energy is
\bee{M6y}
{\cal P}^+\mvec{Z}(t)\ =\ 
\Big(\mvec{a},\mvec{n}, U, \Theta(\mvec{a},\mvec{n}, U,\mvec{\xi}),\mvec{\xi}\Big)(t)\ =\ 
\Big(\mvec{a},\mvec{n}, U,\mvec{\xi}\Big)(t)\ \equiv\ (\mvec{\Omega}^+)(t).
\ee
A physical interpretation is as follows: the contact temperature loses its status as an independent
variable, replacing it by a function of the equilibrium and of the internal variables defined on
the non-equilibrium state space \R{M6y}$_3$ which represents an intermediate concept of the
non-equilibrium state space \R{41}$_1$ and the equilibrium subspace \R{M6}.

The entropy production \R{41}$_2$ becomes on $\mvec{\Omega}^+$ 
\bee{M6w}
\Sigma^+\ =\ \mvec{\beta}^+\cdot\sta{\td}{\mvec{\xi}}
\ee
because $\Theta$ is not a state variable of $\mvec{\Omega}^+$, but a constitutive equation
according to \R{M6y}$_1$.

Analogously to \R{M6a} and \R{M6b}, the time rate of the corresponding non-equilibrium entropy is
%\newpage
\bey\nonumber
\sta{\td}{S}({\cal P}^+\mvec{Z}) &=&
\Big(\frac{1}{\Theta(\mvec{\Omega}^+)}
+\alpha(\mvec{\Omega}^+)\frac{\partial \Theta}{\partial U}\Big)\sta{\td}{U}
-\Big(\frac{{\bf A}(\mvec{\Omega}^+)}{\Theta(\mvec{\Omega}^+)}-\alpha(\mvec{\Omega}^+)\frac{\partial \Theta}{\partial \mvec{a}}
\Big)\cdot\sta{\td}{\mvec{a}}-
\\ \label{M6z}
&&-\Big(\frac{\mvec{\mu}(\mvec{\Omega}^+)}{\Theta(\mvec{\Omega}^+)}-\alpha(\mvec{\Omega}^+)\frac{\partial \Theta}{\partial \mvec{n}}
\Big)\cdot\sta{\td}{\mvec{n}}
+\Big(\mvec{\beta}(\mvec{\Omega}^+)+\alpha(\mvec{\Omega}^+)\frac{\partial \Theta}{\partial \mvec{\xi}}\Big)\cdot\sta{\td}{\mvec{\xi}}\ =\
\\ \label{M22}
&=& \frac{1}{T^+}\Big(\sta{\td}{U}
-{\bf A}^+\cdot\sta{\td}{\mvec{a}}-
\mvec{\mu}^+\cdot\sta{\td}{\mvec{n}}\Big)
+\mvec{\beta}^+\cdot\sta{\td}{\mvec{\xi}}\ =:\ \sta{\td}{S}\!{^+}(U,\mvec{a},\mvec{n},\mvec{\xi}),
\eey
resulting in
\byy{M9y}
\frac{1}{T^+} &:=& \frac{1}{\Theta(\mvec{\Omega}^+)}
+\alpha(\mvec{\Omega}^+)\frac{\partial \Theta}{\partial U}\ =\
\frac{\partial S^+(\Omega^+)}{\partial U},
\\ \label{M10y}
\frac{{\bf A}^+}{T^+} &:=&
\frac{{\bf A}(\mvec{\Omega}^+)}{\Theta(\mvec{\Omega}^+)}
-\alpha(\mvec{\Omega}^+)\frac{\partial \Theta}{\partial \mvec{a}}\ =\ 
-\frac{\partial S^+(\Omega^+)}{\partial\mvec{a}},
\\ \label{M11y}
\frac{\mvec{\mu}^+}{T^+} &:=& \frac{\mvec{\mu}(\mvec{\Omega}^+)}{\Theta(\mvec{\Omega}^+)}-\alpha(\mvec{\Omega}^+)\frac{\partial \Theta}{\partial \mvec{n}}\
=\ -\frac{\partial S^+(\Omega^+)}{\partial\mvec{n}},
\\ \label{M11z}
\mvec{\beta}^+&:=&
\mvec{\beta}(\mvec{\Omega}^+)+\alpha(\mvec{\Omega}^+)\frac{\partial \Theta}{\partial \mvec{\xi}}\ =\ \frac{\partial S^+(\Omega^+)}{\partial\mvec{\xi}}.
\eey

Considering a relaxation process $\cal I$ \R{M13b} or $\cal T$ \R{M13c} along which the equilibrium
variables are fixed, the corresponding time rate of the non-equilibrium entropy \R{M22} is equal to
the entropy production, and the intensive variables \R{M9y} to \R{M11z} depend on the internal
variables. Especially, the temperature \R{M9y} is defined on $\mvec{\Omega}^+$ as a constitutive
equation
\bee{M11z1}
T^+\ =\ \vartheta^+(U,\mvec{a},\mvec{n},\mvec{\xi}).
\ee

Now the following question is investigated: What are the conditions by which the non-equilibrium
temperature \R{M11z1} can be replaced by the thermostatic equilibrium temperature 
\bee{M11z2}
T_0\ =\ \vartheta_0(U,\mvec{a},\mvec{n})
\ee
although non-equilibrium is under consideration ? First of all, the temperature is not an independent
state variable, resulting in 
\bee{M11z2a}
\alpha\ \equiv\ 0
\ee
according to the fourth term in \R{M6a}. If now $\mvec{\beta^+}$ does not depend on the internal
energy, \R{M11z} and \R{M9y} yield
\byy{M11z3} 
\mvec{\beta}^+\ =\ \mvec{\beta}(\mvec{a},\mvec{n},\mvec{\xi})\quad\longrightarrow\quad
\frac{\partial^2 S^+(\Omega^+)}{\partial U\partial\mvec{\xi}}\ =\ \mvec{0},
\\ \label{M11z4}
\longrightarrow\quad
\frac{\partial}{\partial\mvec{\xi}}\frac{1}{T^+}\ =\ \mvec{0}\quad\longrightarrow\quad
T^+\ =\ \theta^+(U,\mvec{a},\mvec{n})\ =: T_0.
\eey
Consequently, the following statement is proved:
\begin{center}
\parbox[b]{12.5cm}{
{\sf Proposition:} If the entropy production on the stage of thermal engineering, described by
internal variables, is independent of the internal energy, the non-equilibrium temperature
does not depend on these internal variables and is therefore
the thermostatic equilibrium temperature although non-equilibrium is in consideration.}
\end{center}

If the entropy production does not depend on the internal energy,
the temperature on the stage of thermal engineering is an equilibrium quantity, whereas
${\bf A}^+$ and $\mvec{\mu}^+$ are not. Consequently, the temperature can be used
as in thermostatics, although the entropy production does not vanish. This fact may be
called {\em contact temperature reduction} ($T^+ \rightarrow T_0$)
which is the reason why in thermal engineering
no problems arise by use of the temperature, why it can be introduced as a "primitive concept"
only depending on the equilibrium variables, and why the contact temperature was not uncovered
earlier.

The thermodynamical stages of Schottky systems introduce an order of their different non-equilibrium
descriptions: the processes of the three stages can be arranged: $\boxplus\Longrightarrow\boxminus$ means,
$\boxplus$ takes more dissipative parts of the process into account than $\boxminus$:
\bee{M24}
\mvec{Z}(t)\quad\Longrightarrow\quad
{\cal P}^+\mvec{Z}(t)
\quad\Longrightarrow\quad{\cal P}\mvec{Z}(t).
\ee

According to \R{M11z3}$_2$, if using the thermostatic temperature, the non-equilibrium
entropy has the shape
\bee{M24a}
S^+(\mvec{\Omega}^+)\ =\ D(U,\mvec{a},\mvec{n}) + B(\mvec{a},\mvec{n},\mvec{\xi}),
\ee
resulting in
\bee{M24b}
\sta{\td}{S}\!{^+}(\mvec{\Omega}^+)\ =\ \frac{\partial D}{\partial U}\sta{\td}{U}
+\frac{\partial(D+B)}{\partial\mvec{a}}\cdot\sta{\td}{\mvec{a}}
+\frac{\partial(D+B)}{\partial\mvec{n}}\cdot\sta{\td}{\mvec{n}}
+\frac{\partial B}{\partial\mvec{\xi}}\cdot\sta{\td}{\mvec{\xi}},
\ee
and as a comparison with \R{M22} shows
\bee{M24c}
\frac{1}{T_0}\ =\ \frac{\partial D}{\partial U},\quad
\frac{\bf A^+}{T_0}\ =\ -\frac{\partial(D+B)}{\partial\mvec{a}},\quad
\frac{\mvec{\mu}^+}{T_0}\ =\ -\frac{\partial(D+B)}{\partial\mvec{n}},\quad
\mvec{\beta}\ =\frac{\partial B}{\partial\mvec{\xi}}.
\ee

Besides the in time local properties which are investigated in the last sections, an in time global
property, the dissipation inequality, is discussed in the next section.

\section{Dissipation Inequality\label{DIIN}}

Starting with the embedding theorem \R{16e} with \R{M6a} and \R{50yy} by taking the entropy
production \R{41}$_2$ into account
\bey\nonumber
&{\cal S}/{\cal R}&\!\!\!\int_{A*}^{B*}
\Big[\sta{\td}{S}(\mvec{Z})-\sta{\td}{S}({\cal P}\mvec{Z})\Big]dt\ =\ 0\ =
\\ \label{62} 
&&=\ {\cal S}/{\cal R}\int_{A*}^{B*}
\Big[\Big(\frac{1}{\Theta}-\frac{1}{T^*}\Big)\sta{\td}{U}
-\Big(\frac{\bf A}{\Theta}-\frac{{\bf A}^*}{T^*}\Big)\cdot\sta{\td}{\mvec{a}}
-\Big(\frac{\mvec{\mu}}{\Theta}-\frac{{\mvec{\mu}}^*}{T^*}\Big)\cdot\sta{\td}{\mvec{n}}
+\Sigma^0\Big]dt,
\eey
the dissipation inequality follows
\bee{M25}
{\cal S}/{\cal R}\int_{A*}^{B*}
\Big[\Big(\frac{1}{\Theta}-\frac{1}{T^*}\Big)\sta{\td}{U}
-\Big(\frac{\bf A}{\Theta}-\frac{{\bf A}^*}{T^*}\Big)\cdot\sta{\td}{\mvec{a}}
-\Big(\frac{\mvec{\mu}}{\Theta}-\frac{{\mvec{\mu}}^*}{T^*}\Big)\cdot\sta{\td}{\mvec{n}}
\Big]dt\ \leq\ 0.
\ee
This inequality does not contain any entropy as well as any entropy production (entropy-free
thermodynamics) \C{MEI72,KE71,MU18a}, the original irreversible process is compared with its
accompanying process. 

Every equilibrium entropy has to be defined uniquely, that means process independently
on the chosen equilibrium subspace. If the considered system is adiabatically unique
\C{MU09}, to each equilibrium entropy non-equilibrium entropies can uniquely be
constructed by use of the entropy production according to \R{M16x}.
By construction, these non-equilibrium entropies satisfy the embedding theorem \R{16e}
which represents a constraint \R{M25} for the contact
quantities. Here, the procedure is the other way round: the non-equilibrium process
induces the thermostatic approximation by reversible accompanying processes.

Besides the dissipation inequality, the efficiency is an other in time global quantity which
is discussed in the next section.

\section{Efficiency of Generalized Cyclic Processes}
\subsection{Generalized cyclic processes}

We consider a cyclic, power producing process of a closed discrete system
which works between two heat reservoirs of constant thermostatic temperatures 
$T^*_H > T^*_L$ ($\boxplus^*$ means reservoir related).
The time dependent contact temperatures of the two contacts between the 
system and the reservoirs are $\Theta_H$ and $\Theta_L$, the net heat exchanges per cycle
through the two contacts $H$ and $L$ are $Q^*_H$ and $Q^*_L$. 
Because a power producing cyclic process is considered, the net heat exchanges satisfy
the following inequalities
\bee{E1}
Q^*_H\ <\ 0,\qquad  Q^*_L\ >\ 0,\qquad 0\ <\ Q^*_L\ <\ -Q^*_H.
\ee

The signs of the heat exchanges through the contacts may depend on time, that means,
the reservoir $L$ does not absorb heat for all times, but there are times for emitting heat.
This behavior is different from that of a Carnot process and typical for {\em generalized cyclic
processes} which satisfy
\byy{E2}
 Q^*_\bx\ =\ \oint\sta{\td}{Q}{^*_\bx}(t)dt\ =\ \oint\Big[\sta{\td}{Q}{^{*+}_\bx}(t)+
\sta{\td}{Q}{^{*-}_\bx}(t)\Big]dt,\qquad\bx\ =\ H,L,
\\ \label{E3}
\sta{\td}{Q}{^{*+}_\bx}\ \geq\ 0,\qquad\sta{\td}{Q}{^{*-}_\bx}\ \leq\ 0.\hspace{3.5cm}
\eey
According to \R{E1}$_3$%                                                      \newpage
\bee{E4}
0\ <\ \oint\Big[\sta{\td}{Q}{^{*+}_L}(t)+\sta{\td}{Q}{^{*-}_L}(t)\Big]dt\ < \
-\oint\Big[\sta{\td}{Q}{^{*+}_H}(t)+\sta{\td}{Q}{^{*-}_H}(t)\Big]dt
\ee
is valid.

\subsection{The mean values}

According to the defining inequality \R{25a}$_1$, we obtain for the closed system
\bee{E5}\ 
\left(\frac{1}{T^*_\bx} - \frac{1}{\Theta_\bx^\dm}\right)\sta{\td}{Q}{_\bx^{*\dm}}\ \geq\ 0,
\qquad\bx\ =\ H,L,\quad\dm\ =\ +,-.
\ee
Using the mean value theorem and \R{E5},
\bee{E6}
\oint\Big(\frac{\sta{\td}{Q}{^{*+}_\bx}}{\Theta_\bx^+}
+\frac{\sta{\td}{Q}{^{*-}_\bx}}{\Theta_\bx^-}\Big)dt\ =\ 
\frac{Q{^{*+}_\bx}}{\LL\Theta_\bx^+\GG}+\frac{Q{^{*-}_\bx}}{\LL\Theta_\bx^-\GG}\ \leq\
\frac{1}{T_\bx^*}\oint\Big(\sta{\td}{Q}{^{*+}_\bx}+\sta{\td}{Q}{^{*-}_\bx}\Big)dt
\ee
is valid. The brackets $\LL\boxplus\GG$ mark the mean value generated by the cyclic process
\bee{E7}
\frac{1}{\LL\Theta_\bx^\dm\,\GG}\ =\ \frac{1}{Q{^{*\dm}_\bx}}
\oint\frac{\sta{\td}{Q}{^{*\dm}_\bx}}{\Theta_\bx^\dm}dt,\qquad\bx\ =\ H,L,\quad\dm\ =\ +,-.
\ee

The extended Clausius inequality of closed systems \R{30b} \C{MU90a,MU83} (one contact,
heat exchange system related)
\bee{E8}
\oint\frac{\sta{\td}{Q}}{T^*}dt\ \leq\
\oint\frac{\sta{\td}{Q}\!(t)}{\Theta(t)}dt\ \leq\ 0
\ee
becomes for the here considered generalized cycle process (two contacts, four reservoir
related heat exchanges)
\bee{E9}
\oint\Big(\frac{\sta{\td}{Q}{^{*+}_H}}{\Theta_H^+}+
\frac{\sta{\td}{Q}{^{*-}_H}}{\Theta_H^-}+
\frac{\sta{\td}{Q}{^{*+}_L}}{\Theta_L^+}+
\frac{\sta{\td}{Q}{^{*-}_L}}{\Theta_L^-}\Big)dt\ \geq\ 0.
\ee

The inequality \R{E6}$_2$ written down for the reservoir $H$ results by use of \R{E4}$_2$ in
\bee{E10}
\frac{Q{^{*+}_H}}{\LL\Theta_H^+\GG}+\frac{Q{^{*-}_H}}{\LL\Theta_H^-\GG}\ \leq\
\frac{1}{T_H^*}\oint\Big(\sta{\td}{Q}{^{*+}_H}+\sta{\td}{Q}{^{*-}_H}\Big)dt\ <\ 0.
\ee
Because of \R{E1}$_1$, the LHS of \R{E10} can be transformed by the mean value procedure
\bee{E11}
\frac{Q{^{*+}_H}}{\LL\Theta_H^+\GG}+\frac{Q{^{*-}_H}}{\LL\Theta_H^-\GG}\ =\
\frac{Q{^{*}_H}}{\LL\Theta_H\GG}\leq\
\frac{1}{T_H^*}\oint\Big(\sta{\td}{Q}{^{*+}_H}+\sta{\td}{Q}{^{*-}_H}\Big)dt\ <\ 0,
\ee
and according to \R{E2}, \R{E11}$_2$ results in
\bee{E12}
\LL\Theta_H\GG\ \leq\ T_H^*.
\ee

Further, \R{E10} in connection with \R{E6}$_1$  and $\bx\equiv H$ results in
\bee{E13}
\oint\Big(\frac{\sta{\td}{Q}{^{*+}_H}}{\Theta_H^+}
+\frac{\sta{\td}{Q}{^{*-}_H}}{\Theta_H^-}\Big)dt\ <\ 0. 
\ee
Consequently, according to \R{E9} and \R{E1}$_2$, the mean value theorem yields
\bee{E14}
0\ <\ \oint\Big(\frac{\sta{\td}{Q}{^{*+}_L}}{\Theta_L^+}+
\frac{\sta{\td}{Q}{^{*-}_L}}{\Theta_L^-}\Big)dt\ =\
\frac{Q{^{*+}_L}}{\LL\Theta_L^+\GG}+\frac{Q{^{*-}_L}}{\LL\Theta_L^-\GG}\ =\
\frac{Q{^{*}_L}}{\LL\Theta_L\GG}
\ee
is valid. The inequality \R{E6}$_2$ written downs for the reservoir $L$ results in
\bee{E15}
0\ <\ \frac{Q{^{*}_L}}{\LL\Theta_L\GG}\ =\ 
\frac{Q{^{*+}_L}}{\LL\Theta_L^+\GG}+\frac{Q{^{*-}_L}}{\LL\Theta_L^-\GG}\ \leq\
\frac{1}{T_L^*}\oint\Big(\sta{\td}{Q}{^{*+}_L}+\sta{\td}{Q}{^{*-}_L}\Big)dt
\ee
from which follows
\bee{E16}
\LL\Theta_L\GG\ \geq\ T_L^*.
\ee

The mean values \R{E12} and \R{E16} follow from \R{E11} and \R{E15}
\bee{E16a}
\LL\Theta_\bx\GG\ =\ Q^*_\bx
\Big(\frac{\LL\Theta_\bx^+\GG\LL\Theta_\bx^-\GG}{Q^{*+}_\bx\LL\Theta_\bx^-\GG+
Q^{*-}_\bx\LL\Theta_\bx^+\GG}\Big).
\ee
If the Schottky system is not a generalized one
\bee{E16b}
Q^{*+}_H\ =\ 0,\quad Q^{*-}_H\ \neq\ 0,\quad Q^{*+}_L\ \neq\ 0,\quad Q^{*-}_L\ =\ 0,
\ee
the equations \R{E11}$_1$ and \R{E15}$_2$ become identities and \R{E12} and \R{E16} are still
valid.

\subsection{Efficiency}

The Carnot efficiency is a statement less applying to generalized cyclic processes because it
is only valid for reversible ordinary cyclic processes. Here, irreversible \U{g}eneralized
\U{c}yclic \U{p}roceses \R{E2} and \R{E3} are considered and the thermostatic temperatures
$T^*_H$ and $T^*_L$ of the controlling reservoirs can be estimated by the inequalities
\R{E12} and \R{E16}  
\bee{E17}
\eta_{CAR}\ =\ 1 - \frac{T^*_L}{T^*_H}\ \geq\ 1 - \frac{\LL\Theta_L\GG}{\LL\Theta_H\GG}\
=: \eta_{gcp}.
\ee
Consequently, the smaller efficiency $\eta_{gcp}$ is a more realistic measure for non-equilibrium processes than the Carnot efficiency $\eta_{CAR}$ which belongs to reversible cyclic processes.

\section{Summary}

Discrete systems in non-equilibrium can be described by use of two different procedures:
the endoreversible treatment \C{MUHO20} or considering the system's internal entropy production
used in this paper. Establishing a non-equilibrium entropy means that also a non-equilibrium
temperature has to be introduced according to \R{B1} or \R{B2}. But then the question arises
why in thermal physics and engineering temperature is introduced as a "primitive concept", that means,
do not ask, if the used temperature is an equilibrium or a non-equilibrium quantity. Clear is, that in
the endoreversible treatment temperature belongs to equilibrium because the system themselves
are in equilibrium. But also if the system is non-equilibrium, temperature is introduced as a
primitive concept or as an arithmetical quantity according to \R{B1}, not knowing, if a thermometer
for a such defined quantity exists. However, in practice thermometers are used measuring
temperatures without clarifying, if these belong to equilibrium or non-equilibrium, and why 
despite this vagueness thermal physics is so extraordinary successful ?  This and other questions are
answered in the following steps:  
\begin{itemize}
\item A directly measurable non-equilibrium temperature \R{25a}$_1$, the contact temperature,
and a non-equilibrium entropy \R{40} are introduced, defined as a state function on the
non-equilibrium state space \R{41} spanned by internal energy, work variables, mole numbers,
contact temperature and internal variables.
\item Because internal energy and contact temperature are independent of each other, both have to
be introduced as independent variables into the non-equilibrium state space.
\item According to the 1st law, the equilibrium subspace is spanned by internal energy, work
variables and mole numbers, the non-equilibrium part by contact temperature and internal variables.
\item Irreversible processes are described by trajectories on the non-equilibrium state space.
\item Every non-equilibrium process generates by projection onto the equilibrium subspace a
reversible accompanying "process" as a trajectory on the equilibrium subspace \R{M5} with the
slaved time as a path parameter.
\item This projection procedure allows to define a non-equilibrium entropy \R{M16x}, if the equilibrium entropy and the entropy production are known.
\item Irreversible processes between two equilibrium states have to be compatible with the
corresponding accompanying processes between the same equilibrium states. This evident
property is garanteed, if the considered system is adiabatically unique and the embedding theorem
\R{16e} is satisfied (Sect.\ref{NEE}).
\item In equilibrium, the non-equilibrium variables --contact temperature and internal variables-- are
functions of the equilibrium variables --internal energy, work variables and mole numbers--.
\item The relaxation of the non-equilibrium variables to their equilibrium values allows to introduce
three so-called thermodynamical stages (Sect.\ref{STTH}):\\
i) The general case of contact quantities: the non-equilibrium variables are not relaxed.\\
ii) The stage of thermal engineering: in general,
the contact temperature depends on the internal variables, but if additionally the entropy production
does not depend on the internal energy (satisfied for nearly all cases), the temperature depends only on the equilibrium variables $T_0=\Theta(U,\mvec{a},\mvec{n})$. 
Thus, its value $T_0$ is the equilibrium temperature which in this case can be
successfully used in non-equilibrium.\\
iii) The endoreversible stage: All non-equilibrium variables are relaxed, the entropy production
is zero, the processes are reversible establishing thermostatics.
\item The global dissipation inequality of entropy-free thermodynamics is revisited (Sect.\ref{DIIN}).
\item The contact temperature makes possible to define an efficiency which is smaller than that
of the reversible Carnot process and which is valid for irreversible and more general cylic processes.
Thus, this efficiency is more realistic than that of Carnot.

\end{itemize}

Beside the introduction to non-equilibrium thermodynamics of Schottky systems, the main result
consists in the answer of the question: Why can an equilibrium temperature successfully be used
in irreversible thermal physics and engineering ?

\end{document}